\documentstyle[12pt,aaspp4]{article}

\lefthead{Kenji Bekki and Masashi Chiba}
\righthead{Formation of the Galactic disk globular clusters}

\begin{document}
\title{Formation of the Galactic disk globular clusters in
early dissipative minor merging}

\author{Kenji Bekki} 
\affil{
School of Physics, University of New South Wales, Sydney 2052, Australia} 

\and

\author{Masashi Chiba}
\affil{
National Astronomical Observatory, Mitaka, Tokyo, 181-8588, Japan}

\begin{abstract}

The origin of metal-rich, highly flattened, and rapidly rotating disk globular
cluster system in the Galaxy is one of longstanding issues in the context of
the Galaxy formation. Our numerical simulations suggest a new ``two-fold''
scenario that the disk globular clusters are firstly formed in the
high-pressure, dense central region of a gas-rich dwarf galaxy, as induced
during the tidal interaction with the pre-existing, young thin disk
of the Galaxy, and then dispersed into the disk region owing to the final
tidal destruction of the merging dwarf. We also demonstrate that spatial
distribution, total number, and metallicity distribution of the clusters
formed in this minor merging depend on the mass ratio of the host to
dwarf galaxy and the orbital configuration of merging.
Based on these results, we discuss whether a minor merging event about 10 Gyr
ago can explain both the Galactic thick disk and the disk globular clusters.
Several other implications for the possible relation between the properties of
disk galaxies and their disk globular clusters are also discussed.

\end{abstract}


\section{Introduction}

The globular cluster system of the Galaxy has been traditionally divided into
two distinct populations: Metal-poor, roughly spherical, slowly
rotating ``halo'' population and metal-rich, flattened, rapidly rotating
``disk'' one (e.g., Zinn 1985, 1991; Armandroff 1989, 1993; van den Bergh 2000).
Recent more detailed studies of the disk globular clusters
have however suggested that the distinction between the two
populations may not be so obvious (e.g., Richtler et al. 1994;
van den Bergh 2000), and furthermore that the metal-rich globular clusters
are associated with the Galactic bulge rather than the disk (e.g., Minniti 1995;
D. Forbes et al. 2001, in preparation).
Physical properties of the disk globular clusters, such as structure,
kinematics,  and metallicity distribution,  are considered to provide valuable
information on the Galactic early dynamical evolution (Zinn 1991).
Owing to the similarity in kinematics, structure, and abundance between
the Galactic thick disk and the disk globular clusters,
the physical relationship between these two Galactic components
have been particularly  discussed (e.g., Armandroff 1993; Norris 1993).

Armandroff (1993) suggested that physical processes responsible for the
formation of the disk globular clusters include: (1) secular scattering of disk
stars and clusters, (2) dynamical heating of the pre-existing thin disk by
merging of one or more dwarf galaxies (Quinn \& Goodman 1986; Quinn, Hernquist,
\& Fullagar 1993), 
(3) gaseous dissipation during tidal encounter between the pre-existing
thin disk and satellites (Ashman \& Zepf 1992), 
and (4) dissipative collapse of the Galactic disk
(Burkert, Truran, \& Hensler 1992). 
However, there have been yet no further theoretical studies which investigate
how the disk globular clusters are formed and whether their
dynamical and chemical properties can be explained by a scenario based on the
above physical process(es). 
Therefore it still remains uncertain which physical process is the most dominant
mechanism for the formation of the disk globular clusters (Armandroff 1993).

The purpose of this paper is to propose a new ``two-fold'' 
mechanism in which the disk globular clusters
are first  formed by dissipative tidal interaction
between a gas-rich dwarf galaxy and the pre-existing, Galactic thin
disk, and then dispersed into the Galactic disk region owing
to tidal stripping of the merging dwarf. 
We show that the high-pressure, dense central region of a merging dwarf,
which is induced by the Galactic tidal force,
can be the site for the formation of the disk globular clusters, because
the high pressure of warm interstellar gas
($P_{\rm g}$ $>$ $10^5$ $k_{\rm B}$: $k_{\rm B}$ is Boltzmann's constant)
can induce  global collapse of giant molecular clouds
and form massive  compact  star clusters
corresponding to globular clusters (Harris \& Pudritz 1994;
Elmegreen \& Efremov 1997).
Based on numerical simulations, we investigate whether this minor merging event
can develop such a high pressure region inside a gas-rich dwarf. 
The present study is the first to demonstrate the detailed predictions of
a certain formation scenario, where we base on a ``two-fold'' mechanism,
to assess whether the model reproduces not only 
metallicity distribution (e.g., Zinn 1985)  but also
orbital properties of the disk globular clusters (e.g., Dinescu et al. 1999).
Therefore we believe that the present study can shed new light on the
formation of the Galactic globular clusters.

\section{Model}

We consider a dissipative minor merger between a large disk with its
dynamical structure similar to that of the Galaxy and a gas-rich dwarf galaxy
represented by a small disk.
Since our numerical methods for modeling chemodynamical evolution of this
dissipative merger, using the TREESPH codes for hydrodynamical evolution,
have already been described  by Bekki \& Shioya (1998) and by Bekki (1995),
we give only a brief review here.
We use the exponential disk model of Fall \& Efstathiou (1980) with
the dark-halo-to-disk mass ratio equal to 4 both for the larger and
smaller disks. The total mass and the size of the larger  (smaller) progenitor
disk are $M_{\rm d}$ ($m_{2} \times M_{\rm d}$, where $m_{2}$ is the mass ratio
of two disks) and $R_{\rm d}$ (${m_{2}}^{1/2} \times R_{\rm d}$), respectively. 
From now on, all the masses and 
lengths are measured in units of $M_{\rm d}$ and  $R_{\rm d}$,
respectively, unless otherwise specified. Velocity and time are 
measured in units of $v$ = $ (GM_{\rm d}/R_{\rm d})^{1/2}$ and
$t_{\rm dyn}$ = $(R_{\rm d}^{3}/GM_{\rm d})^{1/2}$, respectively,
where $G$ is the gravitational constant and assumed to be 1.0
in the present study. If we adopt $M_{\rm d}$ = 6.0 $\times$ $10^{10}$ $
\rm M_{\odot}$ and $R_{\rm d}$ = 17.5 kpc as  fiducial values, then $v$ =
1.21 $\times$ $10^{2}$ km $\rm s^{-1}$   and
$t_{\rm dyn}$ = 1.41 $\times$ $10^{8}$ yr, respectively.
Here, for the sake of clarity in the demonstration of our proposed scenario,
we show the case that {\it only the dwarf galaxy has a gaseous
component}; our experiments indicate that the presence of gas in the larger,
Galactic disk does not affect our conclusion. We set the gas mass fraction of
the smaller disk as 0.1.
An isothermal equation of state is used for the gas 
with temperature of $7.3\times 10^3$ K (corresponding to sound speed 
of 10 km $\rm s^{-1}$).
Guided by the observed metallicity-luminosity relation for dwarf galaxies,
[Fe/H]=$-3.43(\pm 0.14)-0.157(\pm 0.012)\times M_{V}$
(e.g., C\^ot\'e et al. 2000),
where $M_{V}$ is the $V$-band magnitude of a dwarf,
we give initial metallicity to each gaseous particle.
For the model with $m_{2}=0.02$, the corresponding initial metallicity [Fe/H]
is about $-0.82$.
By changing the mass ratio $m_2$, and orbital eccentricity ($e_{\rm orb}$),
pericentric distance ($r_{\rm p}$), orbital inclination of the dwarf's orbit
($\theta$),
we investigate the parameter dependences of physical properties of forming
globular clusters.
We present mainly  the results for our fiducial model with  $m_2$=0.02,
$e_{\rm orb}$=0.5, $r_{\rm p}=0.75R_{\rm d}$, and $\theta = 30^{\circ}$.

We adopt the following two different star formation laws to
convert gaseous particles into either globular clusters or field stars.
For globular clusters,  we adopt the formation model by
Harris \& Pudritz (1994),
in which interstellar gaseous pressure ($P_{g}$) in star forming regions of
a galaxy can drive collapse of pressure confined, 
magnetized self-gravitating  proto-cluster molecular clouds
and form compact clusters, if $P_{g}$ is larger than the surface
pressure ($P_{\rm s}$) of the clouds: 
\begin{equation}
P_{\rm g} \ge  P_{\rm s} \sim 2.0\times 10^5 k_{\rm B}. \;
\end{equation}  
In our simulations, a gas particle is converted into
{\it one}  new cluster  if the gas pressure
is larger than $P_{\rm s}$ = $2.0\times 10^5 k_{\rm B}$. 
Although we cannot investigate the detailed physical processes of
cluster formation in the present {\it global}
(from $\sim$ 100 pc to 10 kpc scale) simulation, we expect that the adopted
phenomenological approach enables us to
identify the plausible formation sites of globular clusters.
For field stars, we adopt the Schmidt law (Schmidt 1959) with the exponent
of 1.5 (Kennicutt 1989). Although this model for star formation is based on
the observed {\it current} star formation law in the Galactic disk and dwarfs,
which may not be simply applied to high redshifts, its details do not affect
the derived results for forming clusters.

\placefigure{fig-1}
\placefigure{fig-2}
\placefigure{fig-3}
\placefigure{fig-4}

\section{Results}

Figure 1 and 2 describe how the Galactic disk globular clusters are developed 
and spatially distributed in the fiducial minor merger model.
As the dwarf passes by the pericenter of its orbit,
the strong tidal field of the Galaxy greatly distorts the dwarf's morphology
to induce efficient energy dissipation in the shocked gaseous regions.
As a result, gas particles having an initially exponential spatial distribution
inside the dwarf are efficiently transferred to the central region of the dwarf,
where the gaseous density and pressure are made high enough to trigger
the globular cluster formation (Figure 2).
Young globular clusters are formed preferentially in the dwarf's central
region with high pressure and density ($T$=0.56 Gyr in Figure 1), 
and they are confined inside the dwarf during the early stage of this
merging event ($T$=2.26).
As the dwarf spirals into the center of the Galaxy owing to dynamical
friction and is consequently destroyed by the strong tidal field
($T$ = 2.82),
the globular clusters are violently  dispersed into   
the Galactic disk region.
The clusters removed from the dwarf appear to retain the orbital angular
momentum of the progenitor dwarf with respect to the Galaxy.
Therefore, the cluster system shows moderate rotation and highly flattened
spatial distribution even after the dwarf destruction
and be identified as ``disk globular clusters'' ($T$ = 3.38, 3.95).
Thus the above simulation demonstrates that the disk globular clusters are
first formed in the central region of the dwarf during the tidal
encounter of the two disks, 
then dispersed around the plane of the Galaxy after
the dwarf's destruction.

Figure 3 shows that the metallicity distribution of the simulated disk
globular clusters has the peak at [Fe/H] $\sim$ $-0.7$ with
the mean metallicity of $-0.58$, that is 1.7 times larger
than the initial metallicity of the dwarf.
This increase in metallicity is due to gaseous chemical evolution
of the starbursting dwarf.
Here, we remark that if the dwarf has already retained globular clusters
which are not formed by the current mechanism and if they are simply dispersed
after this merging event, their typical metallicity is estimated as
[Fe/H]=$-1.45$ using the observed host luminosity vs. cluster
metallicity relation (e.g. C\^ot\'e et al. 2000); such a scenario
without taking into account newly formed clusters fails to reproduce
the observed metallicity of the disk globular clusters.
Thus, our results imply that gaseous dissipation during tidal deformation
of a dwarf, and subsequent star formation and chemical evolution
inside it can play an important role in  the formation
of metal-rich disk globular clusters.
The derived  peak metallicity depends on $m_{2}$ 
such that the peak value is larger for larger $m_{2}$.  
This implies that the dwarf mass should be in a certain range 
in order to reproduce the observed peak value of metallicity
distribution of the Galactic disk globular clusters
($\sim$ $-0.6$, e.g.,  Zinn 1991).

As is shown in Figure 3,  
only a small fraction of the simulated globular clusters (20\%) have
orbital eccentricities ($e$) smaller than 0.3 (i.e., more circular orbits) 
and the mean eccentricity is estimated to be 0.52.
This result is in disagreement with the recent observational result 
by Dinescu et al. (1999) which  suggests that a fairly large  fraction of
the Galactic disk globular clusters show nearly circular orbits ($e$ $<$ 0.3).
A possible reason for this discrepancy is that many of high-$e$ disk globular
clusters obtained here are preferentially destroyed by the Galactic tidal
field, because such clusters pass by the central region of the Galaxy, so that
they are missing at the present epoch. Also, there may exist some observational
bias in the selection of cluster sample; the Dinescu et al. result is
based on a small number of clusters with available proper motions.

The above results are obtained from our fiducial model, but general
properties of globular clusters formed by the current mechanism
are quite diverse depending on the merger parameters.
Figure 4 summarizes the derived diversity.
Firstly, the spatial distribution of disk globular clusters is more centrally
concentrated and more spherical for the model with smaller pericentric distance
($r_{\rm p}$) and larger orbital eccentricity ($e_{\rm orb}$) (panel a).
Secondly, the distribution is less centrally concentrated and more spherical
for the model with  
higher degree of the dwarf's orbital inclination (panel b). 
Thirdly, globular clusters are less efficiently formed 
and rather  centrally concentrated in the retrograde minor mergers (panel c).
This is essentially because tidal disturbance in retrograde merging
is so weak that the developed globular clusters
can not be tidally stripped so readily.
Fourthly, formation efficiency  of disk globular clusters 
and  the mean metallicity are  higher, 
and the spatial distribution is more  centrally concentrated,
for the model with larger $m_{2}$ (panel d) ---
such clusters may correspond to the ``bulge'' globular clusters.
This result furthermore implies that
a disk galaxy with a thicker disk and/or a bigger bulge may
hold a larger number of more metal-rich and more centrally concentrated
disk (or bulge) globular clusters.
This is essentially because  a merger with larger $m_{2}$ can create
both a thicker disk (Bekki \& Shioya 2000) 
and a larger number of more metal-rich globular clusters (this study)
owing to stronger tidal disturbance.
Taking into account the above dependence of the result on model parameters,
we stress that the nature of disk globular clusters in various disk
galaxies may provide us valuable information on their past dynamical histories
of dissipative minor merging.

\section{Discussion and conclusion}

\subsection{Advantages and disadvantages of the present model}

The present numerical study has first demonstrated that 
gaseous pressure of a gas-rich dwarf tidally interacting and merging with
the Galactic disk can become so high ($>$ $10^5$ $k_{\rm B}$) as to
induce global collapse of giant molecular clouds,
which are considered to be progenitor objects of disk globular clusters.
Our numerical simulations have also demonstrated that
if the orbit of the dwarf is nearly coplanar with respect to the Galactic plane,
globular clusters formed by the present mechanism show both
highly flattened density distribution and rapid rotation.
Based on these results, we suggest that the observed structural, kinematical,
and chemical properties of the Galactic disk globular clusters
can be understood in terms of orbital configuration and metal abundance
of gas-rich dwarfs which merge with the Galactic disk. Since galaxy merging
is considered to be important for the formation of thick disks (e.g.,
Quinn et al. 1993) and bulges (e.g., Barnes \& Hernquist 1992), 
the present study implies the existence of some correlation
between the Hubble morphological sequence and physical properties of disk
globular clusters. Thus our scenario has advantages not only in explaining
some of fundamental properties of the Galactic disk globular clusters but also
in providing clues to the better understanding of physical relationship between
formation of disks and globular clusters.

However, it is not so clear whether the presented scenario can explain
the following three aspects of the Galactic disk globular clusters: 
(1) distribution of orbital eccentricities ($e$), (2) three dimensional
spatial distribution, and (3) metallicity distribution. 
The present study predicts that the fraction of nearly circular
orbits ($e$ $<$ 0.3) is only $\sim$ 20 \% whereas 
recent observations (Dinescu et al. 1999) showed it to be 100 \%, although
this estimate is based on only two clusters with [Fe/H] $>$ $-0.8$ in
the Dinescu et al. sample --- the selection of the cluster sample
is biased in terms of availability of proper motions.
Nonetheless, if many of disk globular clusters are characterized by low-$e$
orbits, a fairly large fraction of high-$e$ clusters, formed by the present
mechanism $\sim$ 10 Gyr ago, should have been destructed by some mechanisms
(i.e., some high-$e$ clusters disappear by the present epoch),
so that the present model can be still consistent with the observation
by Dinescu et al. (1999). There are two proposed mechanisms for the
destruction of globular clusters:
(1) bulge shocking (e.g., Aguilar, Hut, \& Ostriker 1988)
and (2) spiral-in to the Galactic center via dynamical friction 
and the resultant destruction of globular cluster and the formation
of the Galactic nucleus (Tremaine, Ostriker,
\& Spitzer 1975).

Aguilar et al. (1988) demonstrated that since the gravitational shocks due to
the central bulge of the Galaxy are very efficient in destroying clusters on
highly radial orbits, clusters passing through the central 2 kpc can be
quickly destroyed by the Galactic central tidal field. The time scale
for a globular cluster to spiral in to the Galactic nucleus owing to
dynamical friction can be estimated (e.g., Bekki \& Couch 2001),
\begin{equation}
t_{\rm fric}= 7.8\times 10^9 \left(\frac {r_{i}}{2{\rm kpc}} \right)^{2}
  \left( \frac{V_{c}}{200 {\rm km \ s^{-1}}} \right)
  \left( \frac{10^6M_{\odot}}{M_{\rm cl}} \right) \ {\rm yr}. \;
\end{equation}
Here we neglect the $\ln \Lambda$ term ($\sim$ 6.6 for a plausible set of
parameters), and $r_{\rm i}$, $V_{\rm c}$, and $M_{\rm cl}$ are the initial
radius from the Galactic center, circular velocity, and mass
of the cluster, respectively. 
Therefore, {\it if a disk globular cluster formed $\sim$ 10 Gyr ago 
can pass through the central 2 kpc or be initially within the central
2 kpc, it cannot be observed at the present epoch owing to tidal destruction
and spiral-in to the Galactic nuclei}. Based on these simple theoretical
arguments, we can investigate what fraction of the simulated disk clusters
with $e$ $>$ 0.3 disappear owing to the above mechanisms.

We choose the simulated clusters with $e$ $>$ 0.3 and investigate
whether or not these can pass through the central 2 kpc of the Galaxy. 
We find that among 56 clusters with $e$ $>$ 0.3, 39 ($\sim$ 70 \%) can pass
through the central 2 kpc within several orbital periods.
This implies that most of the simulated disk clusters cannot survive from
tidal shocking of the Galactic bulge (or spiral-in to the nucleus) and thus
are not identified as globular clusters  at later epochs. 
Therefore, we can expect that the fraction of disk globular clusters
(formed earlier) with $e$ $<$ 0.3 at the present epoch
can be $\sim$ 53 \% in the simulation.
Further measurements of proper motions of many more disk clusters,
by next-generation astrometric satellites such as FAME and GAIA,
are required to assess our prediction.

The observed disk globular clusters are confined inside
about 9 kpc from the Galactic center and about 3 kpc from the Galactic plane
(Zinn 1985). It is an important test whether or not the present model
can reproduce this observational result.
Figure 5 shows that the simulated clusters are confined inside
about 8 kpc from the Galactic center and about 2 kpc from
the Galactic plane, which is basically consistent with observations.
Figure 5 also shows that even if we remove the high-$e$ clusters
(with $e$ $>$ 0.3) having the pericentric distances smaller than 2 kpc
(i.e., those not observed at the present epoch owing to tidal destruction
from the Galactic bulge), the spatial distribution remains almost the same.
Therefore we can say that the present scenario is consistent at least
qualitatively with the observed spatial distribution of the disk globular
clusters.

What we should emphasize here is that observational data
show that many globular clusters are currently within 2 kpc
of the Galactic center.
The existence of these bulge globular clusters appears
to be inconsistent with the results by Aguilar et al. (1988)
and thus implies that the  above discussion
is not so plausible.
We however consider that these bulge clusters may exits owing to the following
reasons: (1) Their orbits are circular with $e$ $<$ 0.3,
(2) these clusters with low $e$ are on the way of disappearance
in a long time scale, (3) these clusters are more tightly bound,
having high densities and small radius,
and (4) these have just arrived at the central 2 kpc
from the outer region of the Galaxy owing to dynamical friction
(i.e., there is not enough time for the Galactic central tidal field
to destroy these clusters). 
Because of the lack of extensive observations on the physical
properties (e.g., orbits, proper motion, and structures) of the central
bulge clusters, it is difficult to determine
why the Galactic bulge clusters are now observed
(i.e., which of the above four explanations is more plausible)
and whether the results by Aguilar et al. (1988)  can be observationally confirmed.
Thus we suggest that the validity of the present scenario,
which requires later destruction of high $e$ 
globular clusters  by the Galactic central tidal field in order to explain
the observed nature  of the disk globular clusters, 
can be assessed by future observations on the bulge globular clusters.

Since the metallicity distribution of globular clusters reflects
the chemical evolution of the Galaxy and the Galactic building blocks
(e.g., van den Bergh 2000), it is also an important test 
for any viable theories  of the Galaxy formation and globular cluster formation.
The present study shows that irrespectively of model parameters,
the metallicity distribution of the simulated clusters is sharply peaked.
On the other hand, observational studies (e.g., Zinn 1985),
showed that the metallicity of disk clusters
with [Fe/H] $>$ $-0.8$  is rather broadly distributed
with the peak metallicity around [Fe/H] $\sim$ $-0.6$.
One possible interpretation of this inconsistency is that owing to some
observational errors in the metallicity scale of globular clusters
(e.g., Sarajedini 1999), the observed metallicity distribution 
of disk globular clusters is smeared out, even if the actual 
distribution is sharply peaked. In order to assess the validity
of this possible interpretation, we examine an average error in
[Fe/H] of disk globular clusters with [Fe/H] $>$ $-0.8$ for the sample of
Zinn (1985) using Table 6 of Zinn \& West (1984), 
and found that it is $\sim$ 0.19 dex.
This result suggests that if we convolve a Gaussian with a dispersion
of 0.2 dex (derived above)  to a metallicity distribution 
with the sharp peak of $-0.6$ dex,
we can anticipate that the resultant metallicity is rather
broadly distributed ranging from $-0.8$ to $-0.4$.
This result implies that the above inconsistency between the present model
and observations is not necessarily a serious problem of the present scenario
of the Galactic disk globular cluster formation.  
In this regard, further, more accurate metallicity measurements of
metal-rich globular clusters are needed to arrive at their true
metallicity distributions (e.g., Carretta et al. 2001; Cohen et al. 1999).

\subsection{An alternative scenario}

We have not presented the effect of a gaseous component in the Galactic disk
on the formation of disk globular clusters, as we focus on
dwarf galaxies as the sites of forming globular clusters.
Dissipative merging has already been suggested to  induce the formation
of globular clusters {\it inside the Galactic gas disk} (Ashman \& Zepf 1992),
and the detailed processes are now being investigated by Bekki \& Chiba
(2001, in preparation). 
This alternative and still attractive scenario might well  have the following
advantages and disadvantages in explaining
structure, kinematics, and chemical properties of disk globular clusters
(Bekki \& Chiba 2001, in preparation).

Firstly, irrespectively of orbital configurations of dissipative merging,
the globular clusters formed inside
the Galactic gas disk can show fairly flattened density distribution
and rapid rotation (for models with a fixed mass ratio of merging two
galaxies).
The formation process which is independent of merger orbital configuration
is attractive in the sense that structural and kinematical properties
of disk globular clusters in external disk galaxies are basically quite
similar to one another.
Secondly, in order to reproduce the observed mean metallicity  
of disk globular clusters, dissipative merging should occur 
{\it only when the Galactic disk has the metallicity of $\sim$ $-0.6$
(i.e., the Galaxy is very young, just after the formation
of the Galactic  stellar halo)}. 
It is not clear at all why such a merging occurs only one time
and at such a preferred epoch. If a couple of dissipative merging events
are involved in the formation of globular clusters, then the resultant
metallicity distribution would show multiple peaks with variously different
values, which is totally inconsistent with observations.

Thirdly, even if dissipative merging that forms disk globular clusters
occurs at a preferred epoch, the metallicity gradient of the gas-rich,
young Galactic disk at this merging epoch should be very different from
that of the present-day Galactic disk.
To be more specific, if the young Galactic disk has the mean metallicity of
[Fe/H] $\sim$ $-0.6$ at the epoch of merging while having
the metallicity gradient being exactly the same as that of typical
present-day disks derived by Zaritsky et al. (1994)
and if globular clusters are formed inside the entire region
of the gas disk, the expected metallicity of disk globular
clusters can range from [Fe/H] = $-1.1$ to $-0.1$.
This appears to be inconsistent with observations and thus the gaseous
metallicity gradient of the Galactic disk at the epoch of merging
should be enough small.

\subsection{A physical relationship between the Galactic thick disk,
halo globulars, and disk ones}

Previous numerical studies have demonstrated that
dynamical heating of the pre-existing Galactic young thin disk by minor
merging of dwarf galaxies can create the thick disk component
(e.g., Quinn \& Goodman 1986; Quinn, Hernquist, \& Fullagar 1993). 
The present results combined with these previous ones
raise  the following two questions: (1) Is just one minor merging event
responsible both for the thick disk formation and for the disk globular
formation? (2) When did such minor merging occur?
Quillen \& Garnett (2000) re-examined the age-velocity dispersion relation
in the solar neighborhood and found that there is an abrupt increase in
the vertical velocity dispersion of the Galactic stellar disk at an age of
9--10 Gyr. They suggested that this abrupt increase can be  explained not
by any gradual and long-term  scattering 
of  disk stars by spiral arms but by minor merging occurred 9--10 Gyr ago.
If this minor merging also results in the formation of disk globular clusters,
the age of the disk globular clusters should be 9--10 Gyr.
Although it is yet uncertain in the age estimate of each disk globular
cluster (e.g., Rosenberg 2000), refining the ages of clusters
will settle the above two issues, or assess the validity of the proposed minor
merging scenario for the formation of the thick disk and disk globular
clusters.

The origin of the observed dichotomy between the Galactic halo and
disk globular clusters is one of unresolved problems of the Galaxy formation.
We propose that the halo globular clusters are formed by multiple
interaction/merging of hierarchically clustered subgalactic clumps or dwarf
galaxies, perhaps seeded by CDM subhalos, {\it prior to disk formation},
whereas the disk globulars are formed by a minor merging of a dwarf with
the Galaxy {\it subsequent to} the first thin disk formation. 
Recent numerical simulations based on a hierarchical clustering scenario
(e.g., Bekki \& Chiba 2001) have demonstrated that metal-poor stars observed
in the Galactic halo were formed initially within subgalactic clumps, which
are developed from initial density fluctuations at high redshifts,
and then dispersed into the Galactic halo region when clumps are disrupted
by tidal stripping. If clumps contain globular clusters, perhaps
formed at the epoch of clump formation, and if these clumps merge violently
with each other at random orientation of orbits, then these globular clusters
will also be widely spread over the halo region during interaction/merging.
The aftermath may represent the halo globular cluster system, having
spherical distribution and very small amount of rotation.

Just after the end of such violent merging events, the first thin disk
component has been at place in the simulated model (Bekki \& Chiba 2001).
Then, some dwarf galaxies, having large orbital angular momentum with respect
to the proto-Galaxy, may arrive at the later epoch.
Such dwarfs eventually interact with the first Galactic thin disk,
accompanying the tidal deformation.
If the initial orbits of these later merging dwarfs are nearly co-planer as
examined here or if the orbits are made more co-planer owing to dynamical
friction from disk stars, then the globular clusters that are formed during
tidal interaction can be dispersed near the disk plane after tidal stripping
from the dwarfs.
Thus we suggest that the dichotomy between the halo and disk globular clusters
may be caused by the difference in the merging epoch of subgalactic clumps
or dwarf galaxies, depending on their initial angular momentum;
cluster populations may be distinguished whether the merging occurred
either before or after the disk formation.

\acknowledgments
We are  grateful to the anonymous referee for valuable comments,
which contribute to improve the present paper.

\clearpage


\begin{figure}
\epsscale{0.45}
\plotone{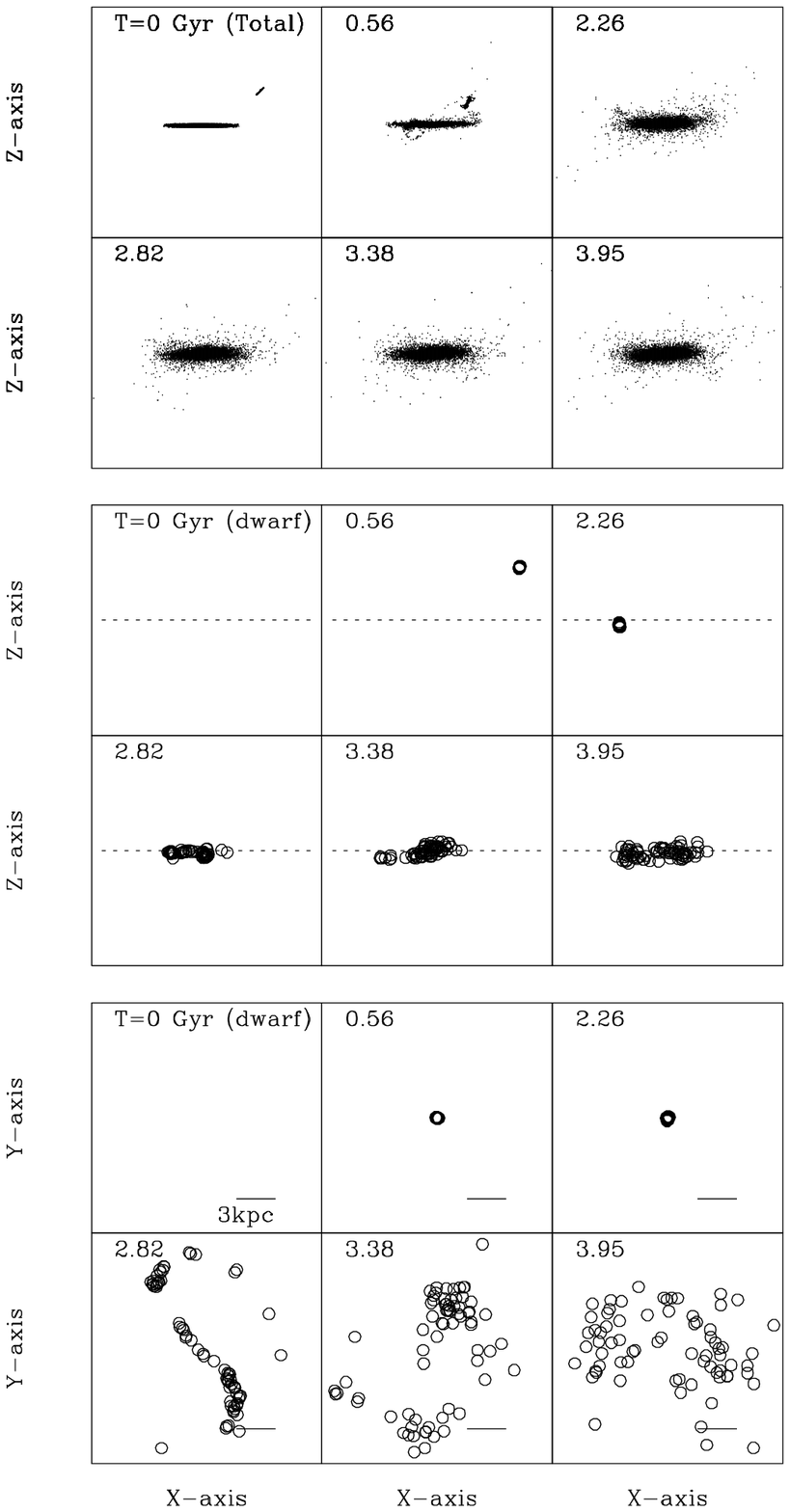}
\caption{
{\it Top six}: Mass distribution of the fiducial merger model 
projected onto  $x$-$z$ plane. All components except dark matter are plotted 
for $T$ = 0, 0.56, 2.26, 2.82, 3.38,  and 3.95 Gyr.
Here $T$ represents the time that has elapsed since the two disks
begin to merge. A frame  measures 105 kpc  on a side for each panel.
{\it Middle six}: Mass distribution of globular clusters (represented
by open circles) projected onto $x$-$z$ plane in the fiducial model. 
In total, 71 globular clusters are formed in this model.
The initial disk position and the size are represented by dotted lines 
and a frame  measures 38.5 kpc on a side for each panel.
{\it Bottom  six}: The same as the middle six but  
projected onto $x$-$y$ plane.
The position of each globular cluster
{\it with respect to the center of the dwarf} is plotted 
and  a frame measures 17.5 kpc on a side.} 
\end{figure}

\begin{figure}
\epsscale{1.0}
\plotone{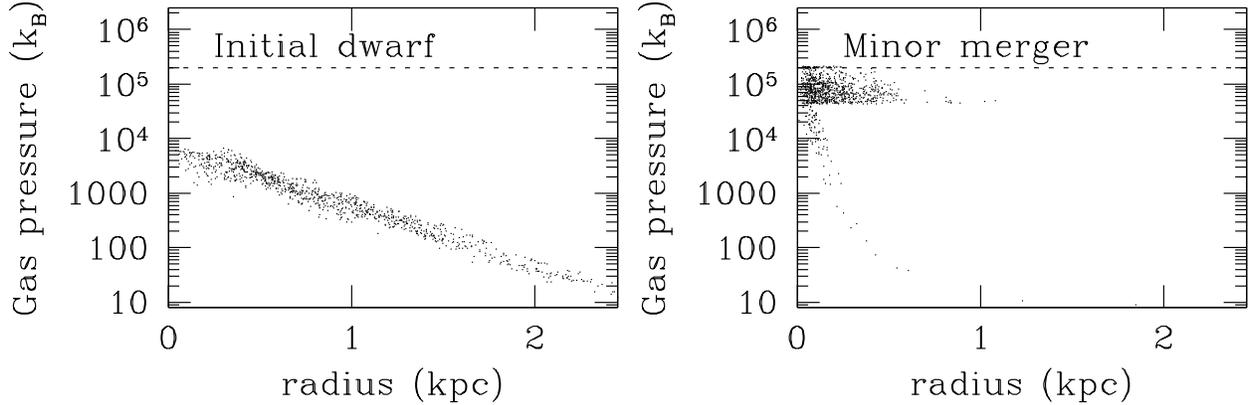}
\caption{
$Left$:  Distribution of gaseous particles  
on a radius-pressure plane  for the initial gas-rich dwarf model.
The dotted line  shows the threshold pressure over which
globular clusters  are assumed to be formed. 
$Right$:  Distribution of  particles
that are $initially$ $gaseous$ $particles$  
on a radius-pressure plane for the fiducial  model 
at $T$ = 0.56 Gyr. 
Here the ``radius'' means the distance
from the center of mass of the dwarf. 
Not only gaseous particles but also new stellar ones  
formed before $T$ = 0.56 Gyr are plotted.
For each new stellar particle, the gaseous pressure
at the epoch when the precursor
gaseous particle is converted into the new stellar one is plotted. 
Accordingly, by comparing the right panel 
with the left one, we can clearly observe
how drastically global dynamical evolution of  dissipative tidal interaction
and minor  merging has increased the gaseous pressure  until $T$ = 0.56 Gyr. 
Note not only that gaseous pressure become rather high during tidal interaction,
but also that the gas is very strongly centrally concentrated 
at this epoch.} 
\end{figure}

\begin{figure}
\epsscale{1.0}
\plotone{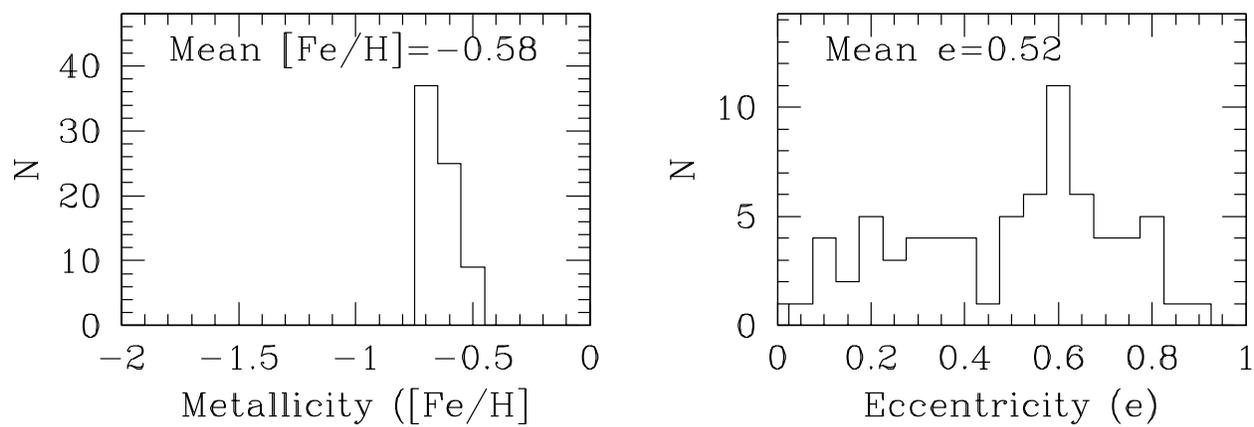}
\caption{
Metallicity distribution (left) and 
the distribution of  orbital eccentricity $e$ (right) for globular clusters
formed in minor merging. Here $e$  for each globular cluster
is defined as $e = (r_{apo}-r_{p})/(r_{apo}+r_{p})$,
where $r_{apo}$ and $r_{p}$ are apogalactic and perigalactic distances
from the center of the simulated Galactic disk, respectively. 
Mean  $e$ and metallicity ([Fe/H]) are given in each panel.} 
\end{figure}

\begin{figure}
\epsscale{1.0}
\plotone{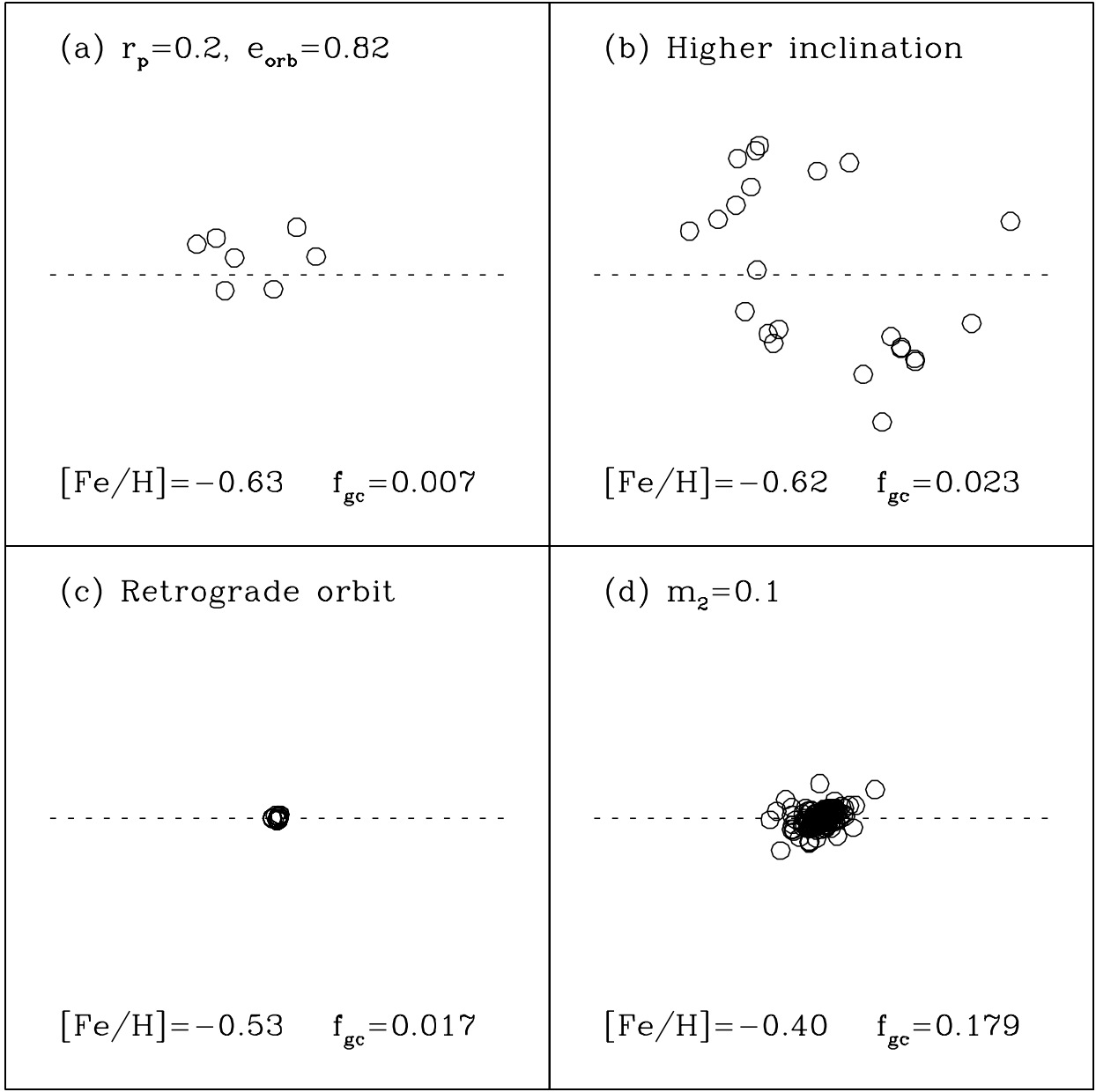}
\caption{
Distribution of the simulated globular clusters (represented by open circles)
projected onto the $x$-$z$ plane for (a) the model with smaller pericentric
distance ($r_{\rm p}$ = 0.2 and $e_{\rm orb}$=0.82,
upper left), (b) high orbital inclination ($\theta$ = ${80}^{\circ}$,
 upper right),
(c) retrograde orbit ($\theta$ = ${150}^{\circ}$, lower left), and
(d) larger mass ratio ($m_{2}$ = 0.1, lower right). 
A dotted line in each panel represents the position and the size of the
initial disk.
The  ratio of the  total number of the developed globular clusters 
to that of initial gas particles (represented by $f_{\rm gc}$)
and the mean metallicity are given in the bottom of the panel for each model.
This $f_{\rm gc}$ can represent the formation efficiency of globular clusters. 
For comparison,  $f_{\rm gc}$ and mean [Fe/H] are estimated
to be 0.071 and $-0.58$, respectively, for the fiducial model.
Each frame measures 42 kpc on a side.}
\end{figure}

\begin{figure}
\epsscale{1.0}
\plotone{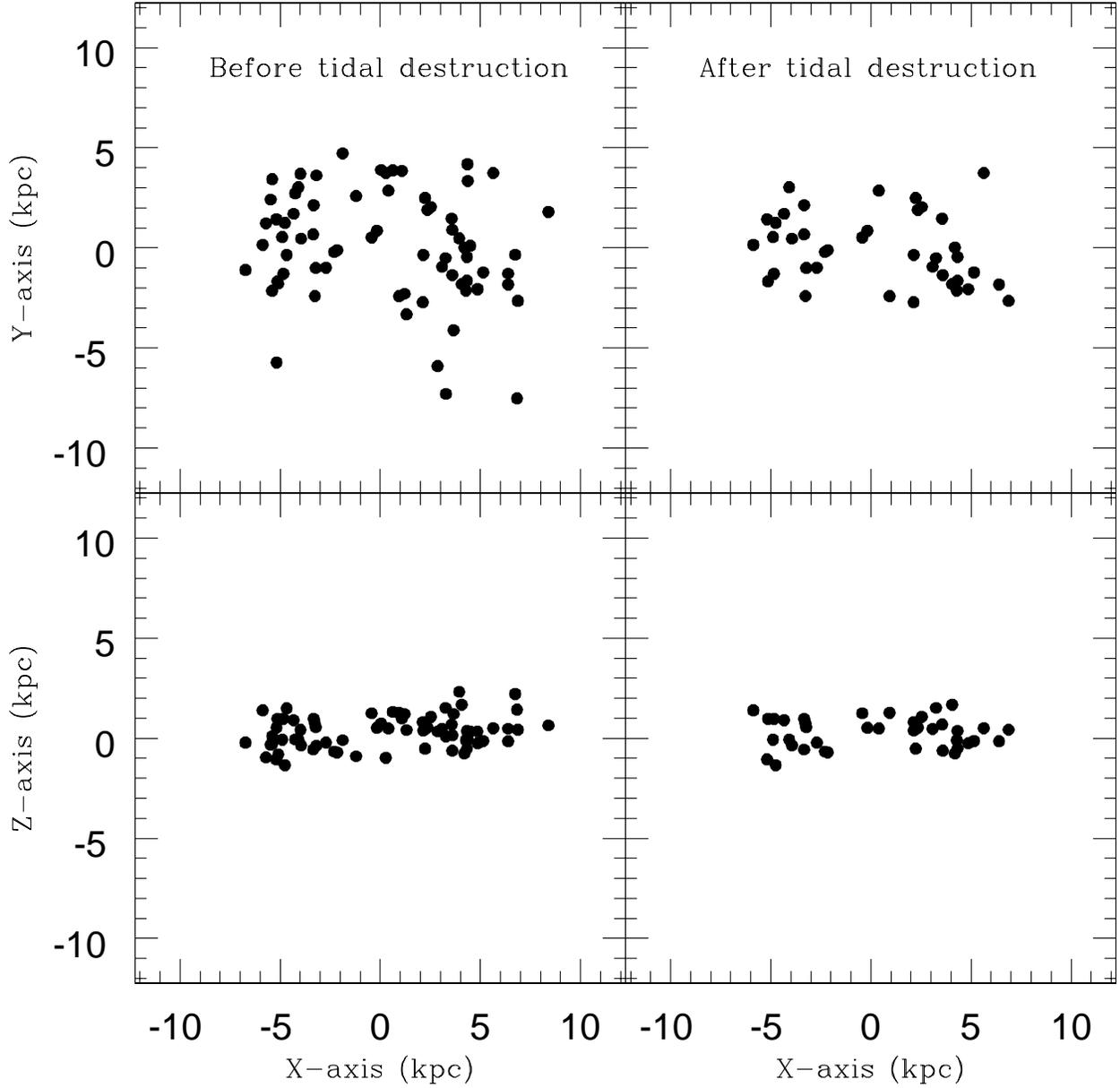}
\caption{
Spatial distribution of the simulated disk globular clusters at $T$ = 3.95 Gyr
projected onto $x$-$y$ plane (upper) and onto $x$-$z$ plane (lower),
for all clusters (left two) and for the survived clusters after the removal
of high-$e$ clusters (with $e$ $>$ 0.3) having the pericentric distances smaller
than 2 kpc (right two) in the fiducial model.
Thus, the left two panels show the spatial distribution of disk globular
clusters just after formation,
whereas the right two ones  show the spatial distribution of clusters
that have survived from tidal destruction (see the text for more details).
Note that even if we consider the tidal destruction effect,
the distribution is not so greatly different between the two cases.}
\end{figure}

\end{document}